\documentstyle[12pt,epsf,psfig,axodraw,rotate]{article} 
\input FEYNMAN 

\newcommand{\nn}{\nonumber}
\newcommand{\be}{\begin{equation}}
\newcommand{\ee}{\end{equation}}
\newcommand{\bea}{\begin{eqnarray}}
\newcommand{\eea}{\end{eqnarray}}
\newcommand{\benn}{\begin{eqnarray*}}
\newcommand{\eenn}{\end{eqnarray*}}

\begin{document}
\newcommand{\nd}[1]{/\hspace{-0.5em} #1}
\begin{titlepage}

\begin{flushright}
NTUA---71/98 \\
OUTP-98--43P \\
cond-mat/9805402 \\
\end{flushright}

\begin{centering}
\vspace{.05in}
{\Large {\bf Probing the 
gauge 
structure 
of high-temperature superconductors \\}}  
 
\vspace{.1in}
{\bf  K. Farakos}$^a$, {\bf G. Koutsoumbas}$^a$, 
and 
{\bf N.E. Mavromatos$^b$}, \\
\vspace{.3in}
{\bf Abstract} \\
\vspace{.05in}
\end{centering}
{\small We suggest that a spin-charge separating 
ansatz, leading to non-Abelian $SU(2) \otimes U_S(1)$ gauge 
symmetries in doped antiferromagnets, proposed earlier as a way
of describing Kosterlitz-Thouless superconducting 
gaps at the nodes of the gap of $d$-wave (high-$T_c$) superconductors,
may also lead to a pseudogap phase, characterised by the formation
of 
(non-superconducting) pairing and the absence of phase coherence. 
The crucial assumption is again the presence of electrically charged 
Dirac fermionic 
excitations (holons) about the points of the (putative)  
fermi surface in the pertinent phase of the superconductor. 
We
present arguments in support of the 
r\^ole of non-perturbative effects (instantons) 
on the onset of the 
pseudogap phase.
As a means of probing such gauge interactions experimentally, 
we  perform
a study of the scaling of the thermal conductivity
with  an externally-applied magnetic field, in certain effective 
models involving gauge and/or 
four-fermion (contact) interactions.}

\vspace{1.2in}

\begin{flushleft} 
$^a$ National Technical University of Athens, Department of Physics,
Zografou Campus 157 80, Athens, Greece, \\
$^b$P.P.A.R.C. Advanced Fellow, 
University of Oxford, Department of (Theoretical) Physics, 
1 Keble Road OX1 3NP, Oxford, U.K. \\
May 1998 \\
\end{flushleft} 

\end{titlepage}

\section{Introduction}

\begin{figure}[htb]
\vspace{0.2in}
\begin{center}
\parbox[c]{4in}{\rotate {\rotate {\rotate 
{\psfig{figure=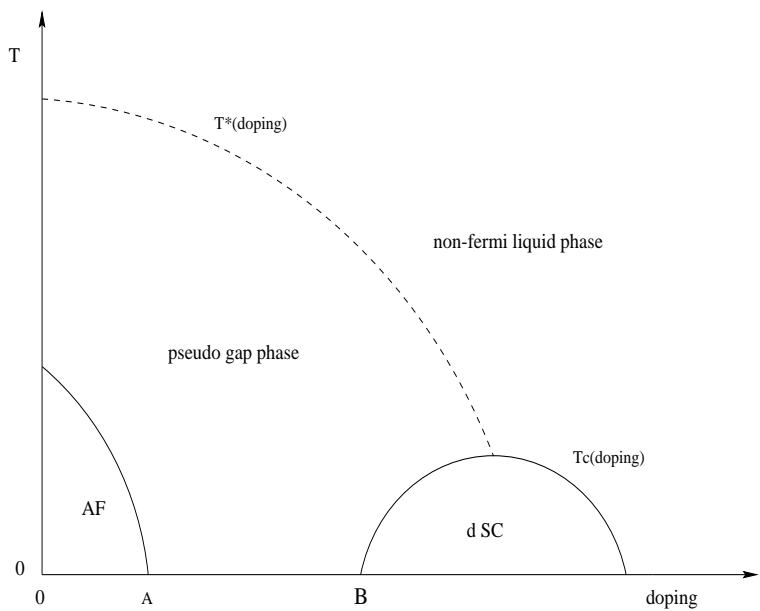,height=10cm,width=8cm}}}}}
\label{pseudogap} 
\end{center}
\caption{{\it The temperature-doping phase diagram summarizes 
the current (experimentally observed) situation in high-temperature 
superconducting cuprates. Notice the existence of an intermediate 
zero-temperature phase, characterised by the existence of preformed pairs, leading to a pseudogap.}}
\end{figure}

Presently, there is 
an abundance of experimental data~\cite{tsuei,underdoped,ong}, 
indicating a very 
rich structure in the 
phase diagrams of the high-temperature superconducting cuprates. 
The pertinent phenomenology may be summarized
by the 
temperature-doping-concentration phase diagram, 
shown in figure \ref{pseudogap}.

The phase diagram shows clearly a very-low (including zero) doping 
antiferromagnetic phase (AF). Above a critical doping 
concetration (point $A$ in fig. \ref{pseudogap}), AF order is destroyed, 
but the interesting issue is the
existence of a phase, named `pseudogap phase', which interpolates 
between the AF and the superconducting 
phases (dSC), the latter being known to be 
of $d$-wave type~\cite{tsuei}.  

It is a
general belief today, supported by 
many experimental results~\cite{underdoped}
on optical conductivity, photo-emission, transport etc., 
that the pseudogap phase is characterized 
by pairing (`preformed pairs'), leading to the existence of a mass 
(pseudo)gap 
in the fermionic spectrum, which however is not accompanied   
by phase coherence. This situation  
is in sharp contradiction 
with 
the
standard BCS theories of superconductivity, according to which 
phase coherence 
appears simultaneously with the gap. 

The pseudogap phase is separated by a critical temperature curve
$T_c$(doping) from the $d$-wave superconducting state, 
the latter being 
characterized by 
a sharp drop in resistivity, and strong type-II superconductivity~\cite{tsuei},
with
the penetration depth for external magnetic fields 
of the order of a few thousands of Angstr\"oms. The pseudogap  phase 
is also separated by another curve $T^*$(doping) from the non-fermi 
liquid normal state phase, where there is no gap,
but there are abnormal properties, such as a  
linear dependence of the electrical resistivity for a wide
range of temperature etc.

The physically challenging situation, therefore, for the high-temperature
materials seems to be not so much their superconducting 
regime~\footnote{Although, the main question why it is $d$-wave rather
than $s$-wave superconductivity, has not, in our opinion, been 
answered yet.}, but rather their behaviour 
in the normal state or under-doped regimes.

Several authors have made various proposals to explain the behaviour 
of the superconductor in those regimes. 
For our purposes here we shall take note of two 
approaches~\cite{marchettilu,fisher}, 
which are closer in spirit to the approach of refs. \cite{dor,fm},
employing effective theories of relativistic fermions as 
the appropriate degrees of freedom for the dynamics of 
the underdoped (or pseudogap ) phase of the cuprates.

In ref. \cite{marchettilu}, the adopted scenario 
is that the fermi surface 
of the underdoped cuprates 
consists of four small pockets, centered around $(\pm \frac{\pi}{2}, 
\pm \frac{\pi}{2})$ in momentum space. The analysis is made 
in the context of a spin-charge separating framework, and 
the charged excitations (holons) 
about the small pockets of the fermi surface 
are treated as relativistic charge excitations. In this latter respect the
assumptions are apparently similar 
with those  in our model~\cite{fm}.  
However, 
the low-energy model used in that work, and the 
nature of the gauge symmetries involved in the 
spin-charge separation ansatz, are different 
from our model. 

In the `nodal liquid' approach of ref. \cite{fisher}, on 
the other hand, the 
starting point is the implementation of quantum disordering 
in the $d$-wave superconductor; 
the relevant excitations 
are relativistic Dirac fermions around the four points constituting
the putative
fermi surface in the underdoped situation~\footnote{These points 
correspond to the 
$d$-wave gap nodes in the superconducting phase.}, which however   
are electrically neutral, and hence, 
from our point of view, they correspond to 
spin degrees of freedom rather than holons. 
The lack of phase coherence in the pseudogap phase is then attributed
in \cite{fisher}, by assumption, to a standard 
Kosterlitz-Thouless transition~\cite{KT} 
of a planar superfluid. 

In the apporach of ref. \cite{fm}, relativistic electrically 
charged 
fermions are employed, 
as the pertinent degrees of freedom, 
describing the excitations about the nodes 
of the energy gap 
of a $d$-wave superconductor.
The analysis  
pointed out the possibility of opening of a gap 
at the nodes of the $d$-wave superconducting gap, below 
a certain temperature, which however was much smaller 
($T_c \le 0.1$ K)
than the 
critical temperature of the $d$-wave superconductor
($T_c^d \sim 100$ K). However, as pointed out in \cite{fmb}, 
upon the influence of an external magnetic field, the 
phenomenon of magnetic catalysis~\cite{miransky} was in operation~\cite{fmb}, 
leading to a scaling of the dynamically induced `nodal 
gap' with the magnetic field, and to an increase of 
the pertinent critical temperature up to 
$\sim 30^o$ K, 
for fields of order $10$ Tesla~\cite{fmb}~\footnote{Note that the 
high-temperature 
superconductors are known to be strongly type II, with a 
London-Meissner penetration depth 
of a few thousands of Angstr\"oms, 
thereby justifying the analysis 
in the presence of an external magnetic field, even in the superconducting state.}. 

In this article, we  
point out that 
the gauge theory of ref. \cite{fm} can also provide 
a possible explanation of the pseudogap  phase
of fig. \ref{pseudogap}. 
We shall also point out 
that it is possible to 
distinguish experimentally  the effects 
of gauge interactions from the ones due to 
four-fermi interactions among the `nodal' holons, such as those 
considered earlier in ref.~\cite{semenoff}. This can be done by 
measuring the thermal conductivity 
in the presence of external magnetic fields, following the experiments of 
\cite{ong}. 
We shall 
give details on the derivation of the scaling of the
thermal conductivity with the externally applied magnetic field, 
based on the phenomenon of magnetic catalysis~\cite{miransky}, 
in {\it both} the 
superconducting and pseudogap phases. 

The structure of the article is as follows: in section 2 we describe
the phase structure of the $SU(2) \otimes U_S(1)$ model, 
with emphasis on the r\^ole of instantons in inducing a pseudogap  phase,
without superconductivity and phase coherence. In section 3 we discuss the behaviour of 
the relativistic charged fermion excitations, argued to describe the 
pertinent excitations about the nodes of the (putative) 
fermi surface in the 
pseudogap phase, in the presence of external magnetic fields.
Then, we study the scaling of the thermal conductivity
with the magnetic field, for various interactions (gauge and four-fermion
type) among the fermions that may characterise the model in various
regimes of the phase diagram. If our model is correct, and 
electrically charged relativistic excitations 
do indeed play a r\^ole in the pseudogap phase, then, such a difference 
in scaling should be 
seen in experiments like those of ref. \cite{ong}, when applied to 
the pseudogap  phase of the cuprates. 
Finally, conclusions and outlook are presented in section 4, where we also 
discuss other ways of detecting the presence of gauge 
interactions in the high-$T_c$ materials. 

\section{Non-perturbative Effects in 
the $SU(2) \otimes U_S(1)$ Gauge Theory and the Pseudogap phase}

The important feature of the 
spin-charge separating ansatz of ref. \cite{fm} was its 
non-Abelian hidden local gauge symmetries, 
$SU(2) \otimes U_S(1)$, 
emanating from a 
`particle-hole' representation of the electronic degrees of
freedom~\cite{affleck}, even in the case of finite doping concentrations. 
The mass gap of the model, and hence the pairing between 
the charged excitations (holons), occurs due to the statistics
changing $U_S(1)$ interactions, which is an exclusive feature 
of the planar geometry of the cuprate materials.
The presence of such interactions in the spin-charge separation
ansatz may be understood~\cite{fm} by means of bosonization 
techniques in three dimensions~\cite{marchetti}. 
The mass generation breaks the $SU(2)$ group down to a $U(1)$ subgroup. 
We now note that the 
$U_S(1)$-induced 
mass gap is characterised by the 
absence of a
local order parameter.
This occurs even in the zero-temperature 
$(2+1)$-dimensional field theory, and it
is a characteristic feature of 
the Kosterlitz-Thouless (KT) nature~\cite{KT} of
gauge theories in $2+1$ dimensions, as argued in \cite{RK,dor}.

When applied to the non-Abelian gauge model of~\cite{fm},
the above symmetry-breaking mechanism
leads to unconventional
KT superconductivity, 
provided that the gauge boson of the unbroken $U(1) \in SU(2)$ 
is {\it massless}. Due to the compactness of the $U(1)$ gauge group, however,
which is a distinctive feature of the non-Abelian gauge group 
of the model \cite{fm}, 
there are non-perturbative effects (instantons), which are 
responsible for giving the gauge boson of 
$U(1)$ a small but finite mass~\cite{polyakov,ahw}. 
This spoils superconductivity,
leaving only a phase, 
characterised by pairing among the 
holons, without the existence of phase coherence. It is one of the points 
of this article to argue that such a phase may provide a 
possible explanation about 
the `pseudogap ' phase of high-temperature 
superconductivity~\cite{underdoped}.

Let us see this in more detail. 
The effective low-energy (continuum) 
theory, describing charged excitations about the nodes of 
a $d$-wave superconducting gap~\cite{fm},  
or the points of the putative fermi surface in the underdoped situation, 
after the integration of magnon (spin) degrees of freedom, reads: 
\be  
{\cal L} = -\frac{1}{4}(F_{\mu\nu})^2 
-\frac{1}{4}({\cal G}_{\mu\nu})^2  +{\overline \Psi}D_\mu\gamma_\mu\Psi +
\kappa \sum_{a=1}^{2} \left( {\overline \Psi}_a \Psi ^a  \right)^2  
\label{su2action}
\ee
\noindent
where 
$\Psi _a$ are the relativistic spinors  
describing the excitations about the nodes of a $d$-wave gap, 
$D_\mu = \partial_\mu -ig_1a_\mu^S-ig_2\sigma^aB_{a,\mu}$,
and $F_{\mu\nu}$, ${\cal G}_{\mu\nu}$ represent the 
field strengths for the $U_S(1)$, $SU(2)$ gauge groups respectively.
For simplicity in this work we work in units $\hbar v_F =1$,
with $v_F$ the fermi velocity for holons, which plays the r\^ole of the 
velocity of light in the relativistic formalism~\footnote{In 
realistic situations~\cite{fmb}, this velocity is smaller than the 
velocity of light $c$ entering the electromagnetic interactions.
However, for the qualitative purposes of this work 
we shall absorb $c$ in the definition of the electric charge $e$.}. 
The four-fermion interactions are assumed attractive,
with $\kappa > 0$. 

\begin{centering}
\begin{figure}[htb]
\vspace{0.2in}
\parbox[c]{4in}{\rotate {\rotate {\rotate 
{\psfig{figure=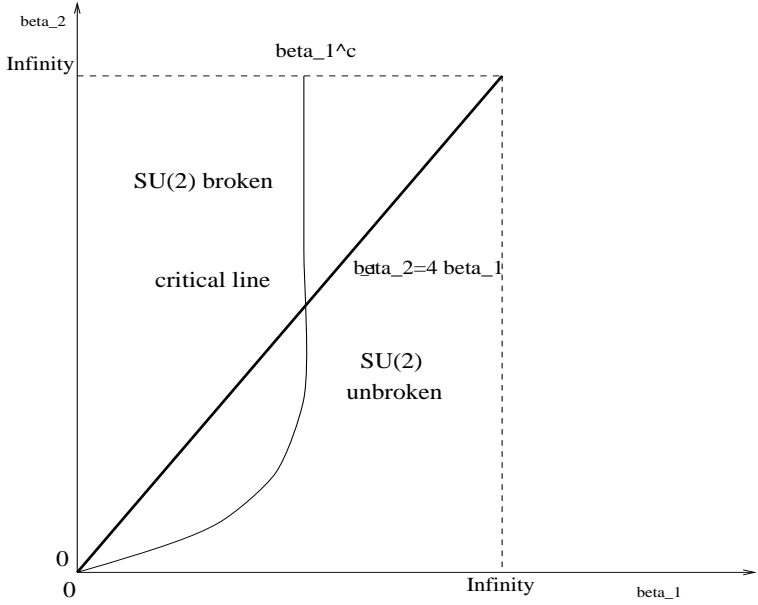,height=10cm,width=8cm}}}}}
\caption{{\it Phase diagram 
for the $SU(2) \times U_S(1)$ gauge theory.
The thick straight line indicates the specific relation 
of the coupling constants in the statistical 
model for doped antiferromagnets used in the present work.}} 
\label{fig21fm}
\end{figure}
\end{centering}

First, let us ignore the four-fermion interactions, assuming them irrelevant.
The gauge $U_S(1)$ 
interaction is capable of 
inducing dynamical opening of a holon gap (pairing) 
if the pertinent coupling constant 
of the statistical model lies inside the $SU(2)$ broken 
regime of the phase diagram of fig. \ref{fig21fm}, 
which is derived in \cite{mavfar}.

An important feature for the onset of a pseudogap  is that the 
non-Abelian gauge group $SU(2)$ breaks down to a compact $U(1)$, generated by the 
$\sigma^3$ Pauli generator of the $SU(2)$ group, as we discussed 
previously~\cite{fm,mavfar}. 
Due to the electric charge of the fermions $\Psi$, the coupling with $U_E(1)$
of electromagnetism causes a spontaneous breaking 
of the electromagnetic symmetry, as can be seen by considering the following matrix element:  
\be
    {\cal S}^a = <B^a_\mu|J_\nu|0>,~a=1,2,3~; \qquad J_\mu ={\overline
\Psi}\gamma _\mu \Psi 
\label{matrix}
\ee
It should be stressed  that as a result 
of the colour group structure only the massless $B^3_\mu $ 
gauge boson of the $SU(2)$ group, corresponding to the $\sigma _3$
generator in two-component notation, contributes to the graph. 
The result is~\cite{RK,dor,fm}:
\be
    {\cal S} = <B^3_\mu|J_\nu|0>=({\rm sgn}{M})\epsilon_{\mu\nu\rho}
\frac{p_\rho}{\sqrt{p_0}} 
\label{matrix2}
\ee
where $M$ is the parity-conserving fermion mass 
(or the holon condensate in the context of the 
doped antiferromagnet). 
As discussed in \cite{dor,RK} 
the $B^3_\mu$ colour component
plays the r\^ole of the Goldstone boson of the (global) $U_E(1)$ 
symmetry~\cite{dor}~\footnote{After coupling with external electromagnetic 
potentials, of course, the (global) fermion-number 
symmetry $U_E(1)$ is gauged and one 
has the Anderson-Higgs phenomenon of symmetry breaking, as explained
in \cite{dor}.}.

If the gauge boson 
$B_\mu^3$ of the 
unbroken $U(1)$ subgroup of $SU(2)$ 
remained {\it massless} exactly, then it 
would be responsible for the 
appearance of a massless pole in the electric current-current 
correlator~\cite{dor}, which would be the 
characteristic feature 
of any superconducting theory. 
The question is whether 
such a pattern of symmetry breaking is capable of explaining the 
non-superconducting intermediate (zero-temperature) phase $AB$ 
of fig. \ref{pseudogap}. At first instance, this seems not possible, in view 
of the appearance of a massless pole in the electric current-current 
correlator~\cite{dor}.

However, in the model of \cite{fm}, 
the compact abelian subgroup $U(1) \in SU(2)$, 
may contain {\it instantons}~\cite{ahw}, which 
in three space-time dimensions 
are like monopoles, and are known to be responsible for giving 
a {\it small} but {\it non-zero mass} to the gauge boson $B_\mu^3$, 
\be 
        m_{B^3} \sim e^{-\frac{1}{2}S_0} 
\label{instmass} 
\ee
where $S_0$ is the one-instanton action, in the dilute gas approximation.
Such a small mass is sufficient to destroy superconductivity of 
the model. This however, from the point 
of view discussed in this section, is a 
welcome result, given that 
it explains naturally the existence of `pre-formed pairs', 
and the opening of a gap (amplitude of the putative order parameter),
but does not lead to superconductivity, in agreement 
with the phenomenology 
of the 
pseudogap  phase.

We now come to the four-fermion interactions. 
The attractive four-fermion 
interactions ($\kappa > 0)$ in (\ref{su2action}) have been taken 
for simplicity to be of the Gross-Neveu type. They arise 
naturally 
in the statistical models 
of interest to us here, 
e.g. as describing the tendency of holes to lie on neighbouring sites 
of the antiferromagnetic lattice~\cite{Sha,dor} or describing 
other contact interactions, appropriate for effective theories
of the Hubbard model~\cite{semenoff}. From the 
pure field-theoretic view point, it is worth mentioning
that such four-fermion interactions in $(2+1)$-dimensions 
become renormalizable (relevant operators) 
in the context of a $1/N$ framework, where $N$ 
is a flavor number for fermions~\cite{gat}. 

\begin{figure}[htb]
\vspace{0.2in}
\begin{center}
\parbox[c]{5in}{\rotate {\rotate {\rotate 
{\psfig{figure=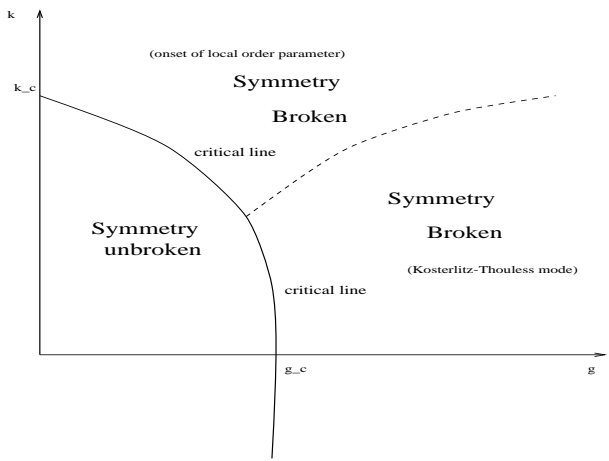,height=8cm,width=6cm}}}}} 
\end{center}
\vspace{1cm}
\caption{{\it A generic phase diagram of the theory with $U_S(1)$ gauge ($g$)
and four-fermion ($\kappa$) 
(Gross-Neveu) interactions. The critical line separates the phase of unbroken symmetry from that of broken symmetry. The symmetry breaking is due to 
the fermion condensate. The dashed line indicates the (possible) onset
of a long-range order (local order parameter) 
due to the dominance of four-fermion interactions. The shape of the line is 
conjectural (at present).}}
\label{fourferd}
\end{figure}

In the presence of $U_S(1)$ and four-fermi interactions (ignoring the 
$SU(2)$ interactions for simplicity), the pertinent phase diagram 
of the model (\ref{su2action}) has been derived, within a $1/N$ 
framework, in ref. \cite{dor}, and is depicted in figure \ref{fourferd}.  

We now remark that in the 
context of the statistical model of ref. \cite{fm}, 
which 
will be the basis 
of our discussion here, the 
various couplings depend explicitly on the doping concentration 
in the sample. In view 
of the phase diagrams of figs. \ref{fig21fm}, \ref{fourferd},
this implies that by tuning the doping concentration appropriately, 
one can induce phase transitions in the model. 

For instance, the 
inverse couplings of the $SU(2)$ and $U_S(1)$ 
gauge groups lie in the thick straight line depicted in fig. \ref{fig21fm}. 
At present, the precise shape of the critical line 
is not known~\cite{mavfar}, 
since it requires the construction of an appropriate  
fermionic algorithm, which will allow for a proper lattice study of the model. 
However, 
the strong coupling analysis of \cite{mavfar} has demonstrated 
that the critical line passes through the origin of the graph. 
For our qualitative purposes, it will be sufficient,
and most likely physically correct,  
to assume an almost vertical shape of the critical 
line from near the upper intersection point 
with the thick straight line 
(away from the origin) till its intersection with the 
($\beta_2=\infty$)-axis.
This implies that 
the critical coupling for mass generation for the coupling $\beta_1$ 
in the statistical model is still given by the  $U_S(1)$ gauge theory 
critical coupling, i.e. 
$\beta_1 < \beta^c_1$ (see figure \ref{pseudogap}).
It is known that $(\beta^c_1)^{-1} \sim \pi^2/32$~\cite{app,kocic}.
Hence, taking into account 
the fact that 
$\beta_1^{-1} \sim \frac{J}{\Lambda}(1-\delta)$ in the model of \cite{fm},
where $\delta$ is the doping concentration in the sample,
and $\Lambda$ is an appropriate ultraviolet cut-off,   
one obtains that 
pairing due to gauge interactions occurs
for:
\be 
      \delta _{AF} < \delta < \delta _c^{(2)} \equiv 1 - \frac{\pi^2}{32}\frac{\Lambda}{J} 
\label{pseudo}
\ee
where 
$\delta _{AF}$ denotes 
the doping concentration at which the AF order is destroyed~\footnote{The 
existence of $\delta_{AF}$ in space-time dimensionality higher than two, 
may de inferred by a renormalization-group analysis~\cite{cp1} 
of the phase 
diagram of the 
magnon ($CP^1$) sector of the model of \cite{fm}. For our purposes here 
we shall not deal with this sector explicitly.}. 
In view of the previous discussion, this region corresponds to the 
pseudogap phase of the diagram of fig. \ref{pseudogap}. 

The onset of ($d$-wave) superconductivity may occur at higher doping 
concentrations, for which 
the attractive 
four-fermi couplings among charged excitations 
in the effective field theory 
become strong enough, so as to overcome the gauge interactions,
and lead to a standard BCS-type pairing among the charged excitations. 

{}From standard arguments~\cite{gat},
we know that pair formation, and hence mass generation, 
in four-fermion Gross-Neveu theories occurs for 
dimensionless inverse couplings, 
$\lambda \equiv \frac{1}{2\kappa \Lambda} $
weaker than a critical value~\footnote{This, of course, pertains to the 
model in the absence of external magnetic fields. The magnetic catalysis 
phenomenon~\cite{miransky} changes the situation~\cite{fmb,ssw}, 
as it may 
lead to four-fermion-interaction induced pairing even for subcritical 
(weak) 
contact interactions. We shall deal with these issues in the 
next section.},
$2/\pi^2$. 
However, the full phase diagram, incorporating 
the $SU(2)$ and $U_S(1)$ couplings as well, 
as appropriate for the model of \cite{fm}, 
will be more complicated. 
However, for our purposes in the present work 
it will be sufficient to 
consider only the effects of the $U_S(1)$ coupling, 
responsible for the mass generation in the model. 

In toy models of doped antiferromagnets~\cite{dorstat,semenoff}, 
the coupling constant $\kappa $ depends on $\delta $: 
\be 
     \kappa \propto \kappa _0 (t', J') \frac{1}{1-\delta} 
\label{kappa0}
\ee
where $\kappa _0 (t',J')$ is an appropriate function 
of the next-to-nearest-neighbor 
hopping element, and Heisenberg exchange energies. 

When combined with the phase diagram of fig. \ref{fourferd}, 
this implies that pairing due to four-fermion interactions 
would occur for doping concentrations in a 
region determined by the critical line of fig. \ref{fourferd}. 
For instance, in the simplifying case when 
the gauge coupling is weak enough,
so that mass generation is due to the four-fermi interactions only, 
this would give: 
\be 
          \delta > \delta_c^{(3)} \equiv 1 - 
\frac{4\Lambda \kappa _0(t',J')}{\pi ^2}
\label{fourfermi}
\ee
By appropriately choosing $\kappa _0 (t',J')$,
it is possible to arrange for a situation like 
the one depicted in 
fig. \ref{pseudogap}, where 
the zero-temperature pseudogap phase interpolates between the AF 
and the standard BCS-type $d$-wave superconducting theory. 

Notice that the dynamical mass generation due to four-fermi couplings
leads to a {\it second order} phase
transition, at zero temperature, 
and hence to phase coherence (local order parameter),
as is standard in BCS-like pairing. This should be contrasted 
with the situation 
in the case of gauge interactions, 
described above, which leads to Kosterlitz-Thouless
type of breaking~\cite{RK,dor,fm} even at zero temperature~\footnote{ The 
finite temperature situation is not discussed here, however one should mention that the coupling of superconducting planes may be necessary in 
maintaining a phase coherence for the four-fermi theory at finite 
temperatures.}.  

Before closing this section we cannot resist in pointing out
that there exist 
some alternative possibilities for the gauge interactions,
which may lead to interesting phase structure in the superconducting 
region of the diagram of fig. \ref{fig21fm}. In one possible scenario, 
as one increases the doping concentration
from the pseudogap phase, 
a point is reached in the $\delta $ axis, where 
there is a special relation among the various 
coupling constants of the effective spin-charge separating theory, 
leading to a $N=1$  supersymmetry~\cite{diamand}~\footnote{This 
supersymmetry
carries non-trivial dynamical information about 
the spin-charge separation mechanism underlying the model,
and hence it is different from the non-dynamical global supersymmetry
algebras, 
at specific points 
of the coupling constants, discovered in \cite{sarkar}.}. 
For example, in the context of models of ref. \cite{dorstat}, such a 
supersymmetric point could be reached 
for doping concentrations 
$t' \sim \sqrt{JJ'} (1-\delta)^{3/2}$. 
Supersymmetry is known~\cite{ahw} to suppress instanton contributions, in the sense that the instanton-induced mass of the $B_\mu^3$ gauge boson is now given by:
\be 
         m_{B^3} \sim e^{-S_0} 
\label{instmass2}
\ee
At such a point, the suppression may be sufficient to 
allow for a gauge-theory induced 
Kosterlitz-Thouless (KT) superconducting gap at the $d$-wave nodes. 
The KT nature of the gap implies that once opened such a gap cannot affect 
the $d$-wave character. This scenario for superconductivity 
has been advocated in ref. \cite{fm}. 

Once a supersymmetric point is reached, 
there is an alternative scenario~\cite{diamand},
which could be in operation in the superconducting 
regime of fig. \ref{fig21fm} for our 
model.
In supersymmetric theories of the type considered 
here and in ref. \cite{diamand},
it is known~\cite{ahw} that  
supersymmetry cannot be broken, due to the fact that the Witten 
index $(-1)^F$, where $F$ is the fermion number, is always non-zero. 
Thus, in supersymmetric theories the presence of instantons 
should give a small mass, if at all, in both the gauge boson and the 
associated gaugino, the latter having been argued in \cite{diamand}
to represent the 
coupling between the superconducting planes. 

However, in three-dimensional supersymmetric gauge theories 
it is possible that 
supersymmetry is broken by having the system in a `false' vacuum~\cite{ahw},
where the gauge boson remains massless, even in the presence of 
non-perturbative configurations, while the gaugino acquires a 
small mass, through non-perturbative effects. 
The lifetime, however, of this false vacuum is very long~\cite{ahw}, 
and hence superconductivity can occur, in the sense that 
the system will remain in the false vacuum 
for a very long period of time, longer than any other time scale 
in the problem. A massless gauge boson would imply superconductivity 
as we discussed above, in the sense of the appearance of a massless 
pole in current-current correlators. In such a scenario,
the false vacuum would occur in the $dSC$ region of the phase diagram,
at the nodes of the $d$-wave gap. The KT nature of the superconductivity 
would imply that the $d$-wave character of the fermi surface 
of the complete statistical system 
is not affected by the opening of a gap at the nodes.

We now point out that, experimentally, 
one can make a distinction 
between a gap induced by the gauge interactions, or by four-fermi 
interactions, as a result of the different scaling of the mass gap with 
an externally applied magnetic field. 
A suggestive experiment along these lines is 
that 
of ref. \cite{ong}, measuring the behaviour of the thermal conductivity,
in both the superconducting and  `pseudogap ' phases.
Theoretically, one needs to develop a formalism for the study 
of the dynamical opening of a gap (fermion condensate) 
in the presence of an external magnetic field, a phenomenon known as 
magnetic catalysis~\cite{miransky}. 
This will be the topic of the subsequent sections.

\section{Interacting Fermions in external magnetic fields} 

\subsection{Review of the Basic Formalism}

In this section we review briefly the theoretical
formalism underlying the behaviour of $(2+1)$-dimensional 
relativistic fermion systems 
under the influence of external magnetic fields~\cite{miransky,shpagin,fkm}. 
One follows 
essentially the method of Schwinger~\cite{schwing}, 
by looking at the coincidence limit of the fermion propagator (in configuration space), ${\rm Lim}_{x \rightarrow y}~S(x,y)|_B$,
in the presence of a constant external field, $B$.
We start first from the free-fermion case, i.e. the case where the
fermions interact only with the external constant field. 
The external gauge potential is given by: 
\be
A^{ext}_\mu = -Bx_2\delta_{\mu1}
\label{extpot}
\ee 
and the Lagrangian is:
\be
  L=\frac{1}{2}{\overline \Psi}(i \gamma^\mu (\partial_\mu - ieA_\mu^{ext}) - m)\Psi  
\label{lagrangian}
\ee
where $m$ is a {\it parity conserving } fermion 
mass, and the $\gamma$-matrices
belong to the reducible  $4\times 4$ representation, appropriate 
for an even number of fermion species~\cite{app,dor,fm}. 

For our phenomenological purposes in this work, we shall assume that 
the mass $m$ is generated {\it dynamically} by 
{\it either} the $U_S(1)$ strong interactions~\cite{fm}
{\it or} the four-fermion contact 
interactions~\cite{miransky,ssw}, in (\ref{su2action}). 
The phenomenon of magnetic catalysis~\cite{miransky} implies that 
there will be a scaling of the fermion condensate with the 
externally applied field, in such a way that 
dynamical mass generation may occur for arbitrarily weak 
couplings, when the external field is 
sufficiently strong. 

Following the proper time formalism of Schwinger~\cite{schwing},
the fermion propagator $S(x,y)|_B=<0|T{\overline \Psi (x)}\Psi (y)|0>|_B$  
in the presence of a constant external magnetic field, $B$,
can be calculated exactly~\cite{miransky}.   
The expansion of 
the (Fourier transform of the) Euclidean 
propagator in terms of the Landau level contributions
is given by~\cite{chodos,miransky}:
\be
 {\tilde S}_E(k)|_B=-i e^{\frac{-k_\perp^2}{e B}} 
\sum_{n=0}^{\infty} (-1)^n \frac{D_n(m,B,k)}{k_3^2+M_n^2(B)},
\ee
with $ M_n^2(B) \equiv m^2+ 2 e B n$, and

\bea
&~&D_n(eB, k)=(m-k_3\gamma_3)
[(1-i\gamma_1\gamma_2{\rm sgn}(eB))
L_n^0(2\frac{{\underline k}_{\perp}^2}{|eB|}) \nn \\
&~&-(1+i\gamma_1\gamma_2{\rm sgn}(eB))
L_{n-1}^0
(2\frac{{\underline k}_{\perp}^2}{|eB|})] \nn \\
&~&+4(k_1\gamma_1 + k_2\gamma_2)
L_{n-1}^1(2\frac{{\underline k}_{\perp}^2}{|eB|}) 
\label{dn}
\eea
with
$L_n \equiv L_n^0,~L_{-1}^1=0,~L_{-1} =0.$
The normalization of the Laguerre polynomials is taken as 
follows~\cite{miransky,ryzhik}:
$   \int _0^\infty e^{-x} L_n (x) L_n (x) dx = 1$.

For our purposes in this work it is useful to quote the result 
of 
the (regularized) magnetically-induced fermion condensate
at zero temperature, after summation of the Landau levels~\cite{fkm},  
in the case of small bare fermion mass $m << \sqrt{eB}$:
\be
\Delta <{\overline \Psi} \Psi>|_{T=0}= \frac{e B}{2 \pi} 
\left(1 + \zeta (\frac{1}{2})\frac{\sqrt{2} m}{\sqrt{eB}}
+ {\cal O}(m^3/(eB)^{\frac{3}{2}}) \right) 
\label{zeroT}
\ee

The finite temperature analysis has been given in 
\cite{fkm}, where we refer the interested reader for details.  
The summation over the Landau poles is possible to be carried out 
analytically 
in some cases. For instance, for a 
strong magnetic field, $B$, 
and low temperatures such that $eB >> T^2>>m^2$, one has: 
$\Delta <{\overline \Psi} \Psi>|_T \simeq 
\frac{e B}{4 \pi} \frac{m}{T} 
-\frac{1.46m\sqrt{eB}}{\sqrt{2}\pi}$, 
using $\zeta (1/2)=-1.46$. This implies the existence of a
{\it critical temperature}, for non-zero $m$, 
\be 
      T_c \simeq \frac{1}{4}\sqrt{eB}
\label{critical}
\ee
above which the condensate vanishes.
The order of magnitude of the temperature
is consistent with the approximations made in the derivation of 
(\ref{critical}), which suggests an important r\^ole for the higher Landau
levels at finite temperatures in inducing a finite critical 
temperature even for the free-fermion case.  
In the context of a possible application of this phenomenon 
to high-temperature superconductors~\cite{fmb}, we note that $\sqrt{eB}$ 
scaling of a critical temperature with the external field intensity 
are reported in the experiments of \cite{ong}.

Notice however, that 
in the presence of extra interactions that may lead to dynamical mass 
generation for fermions, as happens in the $SU(2)\times U_S(1)$ 
model of \cite{fm}, the scaling 
of the critical temperature with $B$ may be different~\cite{fmb}. 
This will be crucial for probing such sturctures experimentally, 
as we discuss in the next subsection. 

Consider first  
the magnetically catalysed condensate in the 
case where there is dynamical mass generation by means of 
$U_S(1)$ gauge interactions~\cite{fmb}.
For strong magnetic fields $\sqrt{eB} >> \alpha$, the lowest-Landau level 
truncation in the Schwinger-Dyson analysis proves sufficient. 
The pertinent low-temperature gap equation reads~\cite{fmb,gusynin}:
\bea
&~& 1 =\alpha \int _0^\infty dx e^{-l^2x/2}\frac{1}{[(\pi T)^2 + x - m^2(T)]^2 
+ (2\pi T)^2 m^2 (T)} \nn \\
&~&[ \frac{(\pi T)^2 + x - m^2 (T) }{m(T)} {\rm tanh}\frac{m(T)}{2 T} + 
\frac{(\pi T)^2 + m^2 - x}{\sqrt{x}} {\rm coth}\frac{\sqrt{x}}{2T}]
\label{gapeq}
\eea
where $\alpha$ is the (dimensionful) fine-structure constant 
of the $U_S(1)$ theory, $T$ is the temperature, and $l$ is proportional to the 
magnetic length, $l^2 \equiv 1/eB$. 
By following the same semi-analytic procedure as 
in ref. \cite{fmb}, it is easy to see that for strong enough fields, and 
low enough temperatures we are interested in here, 
$T << m$, 
the gap equation simplifies to: 
\bea
&~&m/\alpha \simeq \left( {\rm tanh}(m/2T) - (2T/m) \right) 
e^{-\frac{m^2}{2eB}} \int _{-1}^0 \frac{dz}{z} e^{-\frac{m^2}{2eB}z}
+ \nn \\
&~&[\left( {\rm tanh}(m/2T) - (2T/m) \right)e^{-\frac{m^2}{2eB}} + 2T/m ]
\int _0^\infty \frac{dz}{z} e^{-z} 
\label{simple}
\eea
Regularizing the infinities of the integrals around $z \sim 0$, as in 
ref. \cite{fmb}, 
in the limit $T/m \rightarrow 0$, 
one obtains:
\be 
m \sim \alpha [{\rm tanh}\frac{m}{2T} - \frac{2T}{m}]e^{-\frac{m^2}{2eB}}
{\rm ln}\left(\frac{2eB}{m^2}\right)
\label{mimproved}
\ee
which for strong enough magnetic fields and low temperatures, 
leads to the 
following 
(approximate) expression for the (temperature dependent) mass gap:
\be
      m_b^{g} (B,T)= m_B(B,0) -|{\cal O}(T/m)| =
 C \alpha {\rm ln}\frac{\sqrt{2eB}}{\alpha}
-|{\cal O}(T/m)| 
\label{gaugeB}
\ee
where $m_B(B,0)$ is the zero-temperature solution~\cite{fmb,shpagin}.
A numerical estimate of $C\sim \sqrt{2}$ has been made in ref. \cite{fmb}
by a zero-temperature analysis. 
The solution (\ref{gaugeB}) 
is consistent with the 
the existence of a critical temperature $T_c$, which has been estimated in 
\cite{fmb} to be:  
\be
T_c \propto \alpha {\rm ln}\frac{\sqrt{2eB}}{\alpha} 
\label{tempgauge}
\ee
The formula (\ref{gaugeB}) suggests that 
for the low-temperature 
regime $T << T_c$, and strong magnetic fields,   
the replacement $m_g(T,B) \sim m(0,B)$ 
proves sufficient for 
an estimate of the leading scaling behaviour 
of the thermal conductivity with $B$. 
 
We next remark that 
in the case of the subcritical Gross-Neveu model~\cite{miransky,ssw}
the corresponding 
zero-temperature mass gap 
for strong magnetic fields scales as: 
\be 
     m_b^{NJL} \sim 0.45 \sqrt{eB} + \dots 
\label{njlg}
\ee
where the $\dots $ indicate subleading terms, proportional to the 
deviation of the four-fermion 
coupling from its critical value~\cite{gat}. At present the analysis for
the supercritical four-fermion model is not available, but it presents 
no conceptual issues. 
As in the gauge case, the gap (\ref{njlg}) 
will be used for an estimate of the low-temperature 
thermal conductivity scaling. 

As we shall show below, in this region of the parameters, the scaling of the 
thermal conductivity in the case of the $U_S(1)$-induced gap~\cite{fm} is 
different from that of the free-fermion~\cite{fkm} and 
four-fermi Gross-Neveu 
models~\cite{miransky,ssw}.

\subsection{Thermal Conductivity}

The expression for the thermal conductivity can
be given in terms of the energy-current (momentum) correlation
function~\cite{Kubo}:
\be
{\kappa}^{el} (\omega ) =\frac{\beta}{V}
\int _0^\infty dt \int _0^\beta d\lambda 
{\rm Tr} \{ \rho_0 P^i (0) P^j (t + i\lambda )\}e^{-i\omega t} 
\label{kuboform}
\ee
where $V$ is the volume of the system, $\beta=1/T$ is the inverse temperature
(in units $k_B=1$), and 
$P(t)=e^{iHt}P(0)e^{-iHt}$, 
$P^i(0)=\frac{i}{2}\int d^2x \
({\overline \Psi}\gamma ^0 
\partial^i \Psi - \partial^i {\overline \Psi}\gamma ^0 \Psi)$,
with $H$ the Hamiltonian; $\rho_0=\frac{1}{Z}{\rm Tr}e^{-\beta H}$ 
is the Gibbs equilibrium distribution, $Z$ is the partition function
of the system,
and the trace ${\rm Tr}$ is taken over all physical states.

Following ref. \cite{ssw}, one can compute the 
static thermal conductivity of an isotropic system,
which will be of interest to us here:
\be
\kappa^{el}=\frac{1}{4T}\lim_{\omega\to 0}\frac{1}{\omega}
\left(G(p=0,i\nu_m=\omega+i0^{+})-G(p=0,i\nu_m=\omega-i0^{+})\right).
\label{cond}
\ee
where 
the thermal Green function can be calculated perturbatively
in the theory~\cite{ssw}:
\begin{equation}
G(p=0,i\nu_m)=iT\sum_{n=-\infty}^{+\infty}
\int \frac{d^2 k}{(2\pi)^2} k^2
\mbox{tr}\left(\gamma^{0}S(i\omega_{n},k)
\gamma^{0}S(i\omega_{n}+i\nu_m,k)\right),
\label{green}
\end{equation}

We follow ref. \cite{ssw} and 
use the following spectral representation for the fermion thermal Green
function:
$S(i\omega_{n},k)=\int\limits_{-\infty}^{+\infty}\frac{d\omega}{2\pi}
\frac{a(\omega,k)}{i\omega_{n}-\omega}$, where 
$a(\omega,k)=2Im S(i\omega_{n}=\omega-i0^{+},k)$. 
Summation over the Landau levels 
implies: 
\bea
&~&a(\omega, k)=2\pi {\rm sgn}(\omega ) \sum_{n=0}^\infty (-1)^n \delta (\omega ^2 - M_n^2(eB))
(m + \omega \gamma _0) e^{-\frac{{\underline k}_{\perp}^2}{eB}} 
\otimes \nn \\
&~&[(1-i\gamma_1\gamma_2{\rm sgn}(eB))
L_n(2\frac{{\underline k}_{\perp}^2}{|eB|})- \nn \\
&~&(1+i\gamma_1\gamma_2{\rm sgn}(eB))
L_{n-1}(2\frac{{\underline k}_{\perp}^2}{|eB|})] + \nn \\
&~&8\pi {\rm sgn} (\omega )\sum_{n=0}^\infty (-1)^n 
\delta (\omega^2 - M_n(eB)^2)e^{-\frac{{\underline k}_{\perp}^2}{eB}} 
(-i\gamma_1 k_1)L_{n-1}^1 (2\frac{{\underline k}_{\perp}^2}{|eB|})
+ \nn \\
&~&4 \sum_{n=0}^\infty (-1)^n P(\frac{1}{\omega^2 - M_n^2 (eB)})
e^{-\frac{{\underline k}_{\perp}^2}{eB}}(-k_2\gamma_2) 
L_{n-1}^1 (2\frac{{\underline k}_{\perp}^2}{|eB|})
\label{sdetails}
\eea

Substituting this  spectral representation into the spectral
representation for $S(i \omega_n,k),$ using the result into 
Eq.~(\ref{green}), 
and performing the sum over the Matsubara frequencies, 
we arrive at the following expression for the 
thermal conductivity~\cite{ssw}: 
\be
\kappa^{el}=\frac{1}{16T^2}
\int \frac{d^2 k}{(2\pi)^2}\int \frac{d\omega}{2\pi}
\frac{k^2}{ \cosh^{2}\left(\frac{\omega}{2T}\right) }
\mbox{tr} \left(\gamma^{0}a(\omega,k)
\gamma^{0}a(\omega,k)\right).
\label{tc}
\ee
Substituting (\ref{sdetails}) in (\ref{tc}),  
we obtain after some tedious algebra: 
\be
\kappa^{el} =\frac{3}{16}\frac{(eB)^2}{T^2}\delta (0) \sum_{n=1}^\infty
\frac{n}{{\rm cosh}^2(\frac{M_n}{2T})} + \frac{(eB)^2}{16T^2}\delta (0)
\frac{1}{{\rm cosh}^2(\frac{m}{2T})}
\label{finalkappa}
\ee
Above we have used the following mathematical identities for 
the $\delta$-functions: 
\bea
&~&\delta\left(\omega^2-M_1^2 \right)
\delta\left(\omega^2-M_2^2 \right) = \nn \\
&~&\frac{1}{4M_1M_2}\left(
\delta\left(\omega+M_1\right)\delta\left(\omega+M_2\right)+ 
\delta\left(\omega-M_1\right)\delta\left(\omega-M_2\right)\right)
\label{dindet}
\eea
by means of which 
the double summation over Landau levels is reduced to a single one. 
We have also made use of the identities of the Laguerre 
polynomials~\cite{ryzhik}: 
\be
\int _0^\infty e^{-x} x L_n (x) L_n (x) dx = 2n + 1,~
 \int _0^\infty e^{-x} x^2 L_n^1 (x) L_n^1 (x) dx = 2(n +1)^2 ~.
\label{lident}
\ee

Now we are in a position to state the main result of the present work, 
namely to 
estimate the scaling of $\kappa ^{el}$ with the magnetic field, for the 
two cases of dynamical mass generation, induced by gauge~\cite{fm,shpagin} 
or four-fermion fields~\cite{miransky,ssw}, the latter interaction being
taken to be of the Gross-Neveu type for simplicity. 
Following \cite{ssw}, 
we also replace $\delta (0)$ in the expression for the thermal conductivity 
(\ref{finalkappa}) 
by the inverse of the width $\Gamma$ for the quasiparticle states, 
$1/\Gamma $, where  
$\Gamma$ is assumed considerably 
smaller than the dynamical mass, $\Gamma << m$, 
in order to preserve the validity of the symmetry
breaking analysis due to the magnetic catalysis.

\begin{figure}[htb]
\vspace{0.2in}
\begin{center}
\parbox[c]{4in}{\rotate {\rotate {\rotate 
{\psfig{figure=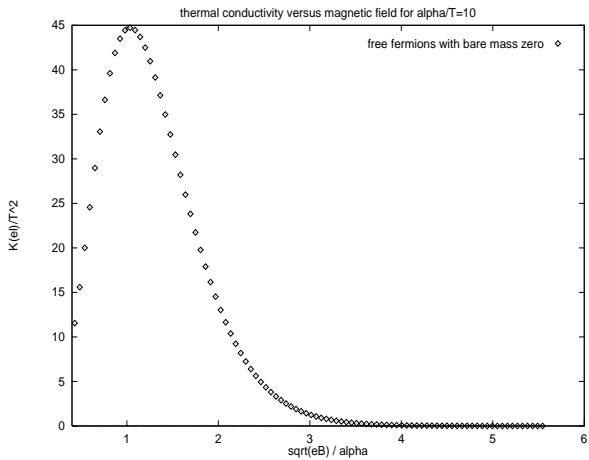,height=8cm,width=6cm}}}}}
\label{free} 
\end{center}
\caption{{\it Scaling of the thermal conductivity for relativistic
electrically charged fermions with a magnetic field, in the free case, 
with vanishing bare mass of the fermions,
as compared to the $\sqrt{eB}$. The temperature and mass 
scales are all given in units
of the $U_S(1)$ gauge structure constant, to make easier the comparison 
with the gauge case.}}
\end{figure}

\begin{figure}[htb]
\vspace{0.2in}
\begin{center}
\parbox[c]{4in}{\rotate {\rotate {\rotate 
{\psfig{figure=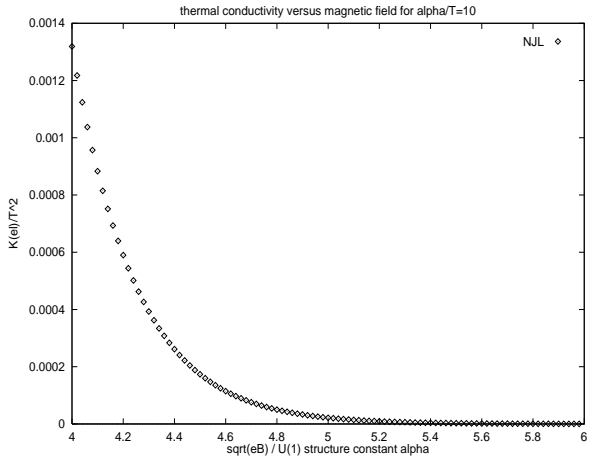,height=8cm,width=6cm}}}}}
\label{tcffermi} 
\end{center}
\caption{{\it Scaling of the thermal conductivity for relativistic
electrically charged fermions with a magnetic field, in 
the case of sub-critical Gross-Neveu four-fermion 
interactions. 
The diagram 
pertains to 
strong magnetic fields and low-temperatures, so that the 
lowest-Landau level analysis is reliable.  
The temperature and mass scales are all given in units
of the $U_S(1)$ gauge structure constant.}}

\end{figure}

\begin{figure}[htb]
\vspace{0.2in}
\begin{center}
\parbox[c]{4in}{\rotate {\rotate {\rotate 
{\psfig{figure=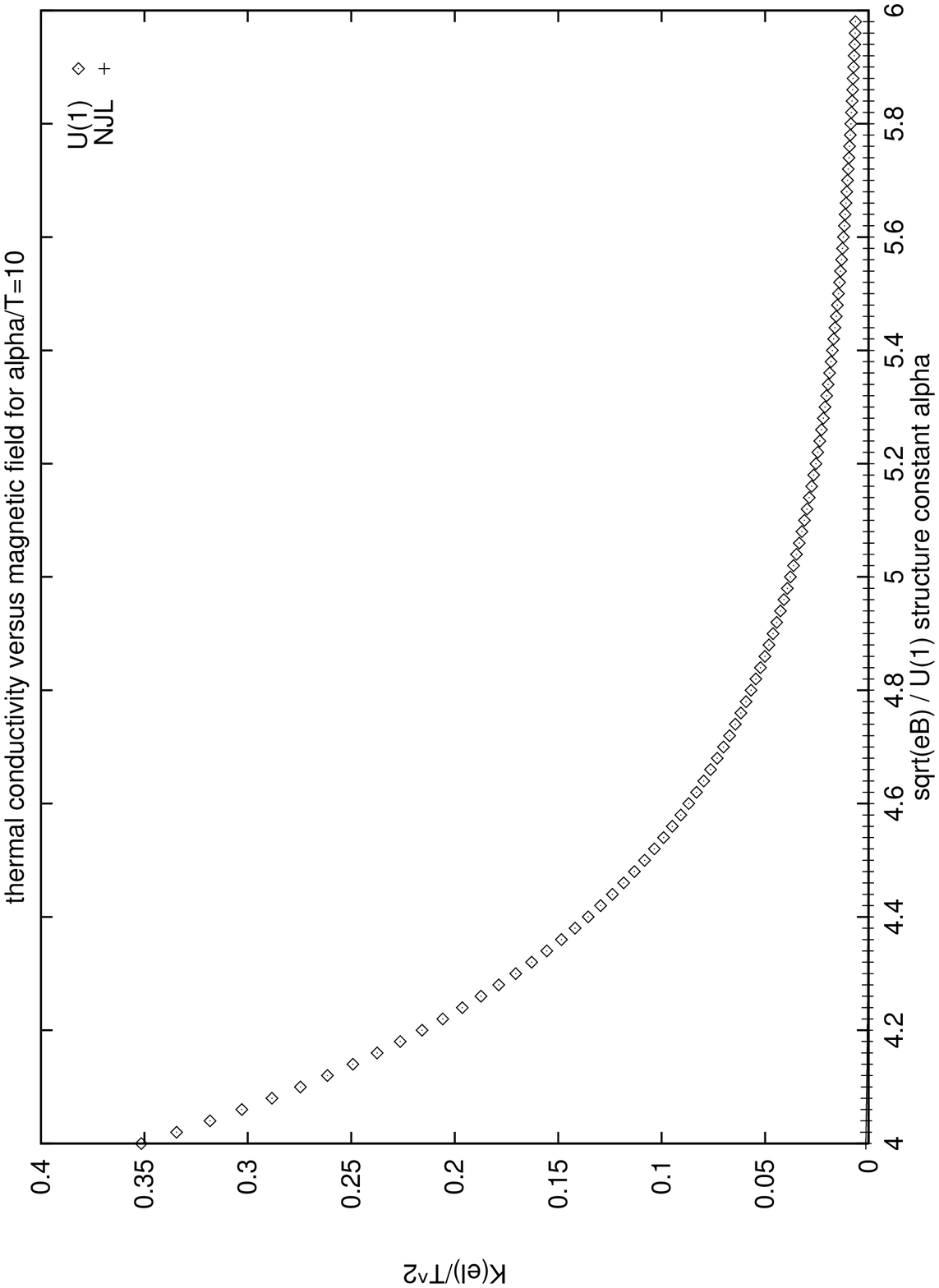,height=8cm,width=6cm}}}}}
\label{tcgauge} 
\end{center}
\caption{{\it Scaling of the thermal conductivity for relativistic
electrically charged fermions with a magnetic field, in 
the $U_S(1)$-gauge-interaction case. The diagram 
pertains to 
strong magnetic fields and low-temperatures, so that the 
lowest-Landau level analysis is reliable. 
For comparison we have also included the result 
for the Gross-Neveu model (see previous figure).  
The temperature and mass scales are all given in units
of the $U_S(1)$ gauge structure constant. }}

\end{figure}

The pertinent results are summarised in figs. 4, 5 and 6, 
for the cases of (a) free relativistic fermions in an external field, 
with vanishing bare mass as compared to $\sqrt{eB}$~\cite{fkm}, 
(b) sub-critical 
Gross-Neveu model~\cite{ssw}, and 
(c) $U_S(1)$ interactions~\cite{fm,dor}, 
respectively. The figures show the thermal conductivity 
versus the applied magnetic field, and - in order to facilitate 
the comparison -  
we have expressed the various scales in units of the 
dimensionful structure constant $\alpha $ of the $U_S(1)$ interactions
in all three cases.

The analysis in all cases has been made at a fixed low temperature,
which allows some analytic results to be derived from 
the respective formalism of refs. \cite{miransky,fkm,fmb,ssw}. 
From the results of the previous subsection, it is clear that 
the 
critical temperatures $T_c$ (\ref{critical}), (\ref{tempgauge}) 
for the magnetic catalysis phenomenon~\cite{fmb,fkm}
scales with the magnetic field appropriately in each case. 
Thus, 
for fixed $T < T_c$, as is the case in the figures, 
this implies a minimum $B$ above which the magnetic 
catalysis phenomenon occurs. This has been taken properly into account 
in drawing the figures.

In the free fermion case, summation over {\it all Landau levels} 
in the expression for the thermal conductivity 
is possible~\cite{fkm}. 
In the case of vanishing bare mass, as compared to $\sqrt{eB}$, 
the resulting magnetically-induced 
condensate at any finite temperature is 
given by the zero-temperature 
result (\ref{zeroT})~\cite{fkm}, provided that $T < T_c$, 
with $T_c$ given by (\ref{critical}). 
The thermal conductivity 
shows a peak. 
If one takes the behaviour of fig. 4 literally, 
then, the presence of a low-magnetic field region, 
where the conductivity increases with the magnetic field, 
contradicts the experimental situation 
of ref. \cite{ong} in the superconducting phase of the high-$T_c$ 
cuprates, where the conductivity reduces with the magnetic field
for low fields. This seems to 
point towards the fact that other interactions 
set in in such regimes. 

In the subcritical 
four-fermion (Gross-Neveu ) case~\cite{ssw}, 
the scaling for strong magnetic fields, 
is similar to the free fermion case. 

In the gauge case, on the other hand, 
the logarithmic scaling 
of the mass gap (\ref{gaugeB}) 
with the external field is a distinctive feature,
and leads to a different scaling 
for the thermal conductivity, 
as compared to the previous cases. This 
is clear from the figs. 5,6.
We also note that 
for the regime of the temperatures considered in fig. 6, 
the condition $T < T_c$, or equivalent $B$ larger than a minimum value,
implies that the peak which occurs 
in the thermal conductivity versus $B$ (as in the case of fig. 4), 
even in the gauge case, lies outside 
the allowed region of $B$. 

The interesting question is whether the gauge interactions 
lead to a scaling of the thermal conductivity in the low-field region 
which agrees with the results of ref. \cite{ong}. 
Unfortunately, at present, analytical results are known only for 
the case of 
very strong fields, where only the lowest Landau level contribution is taken 
into account. Further studies along this direction are in progress. 

However, even at this preliminary stage, our analysis above has demonstrated 
that in principle, thermal conductivity experiments, under the influence 
of external magnetic fields,  
may be useful probes of 
gauge structures in 
the dynamics of high-temperature superconductors, provided one 
performs the experiments in different regions of the 
phase diagram of fig. \ref{pseudogap}. If there is a regime of doping 
for which gauge interactions set in, 
then there should be a difference in scaling of the thermal conductivity 
with the external field, as compared to the regime where the standard 
BCS type four-fermion interactions among the quasiparticles are present. 

It is of course, understood, that our preliminary analysis above pertains
only to the thermal conductivity of relativistic,
electrically charged,  
`nodal' quasi-particle 
excitations, which have been argued to 
play a r\^ole in {\it both} the pseudogap  
as well as the $d$-wave superconducting phases~\cite{fmb}. 
In actual experiments, the total thermal conductivity 
receives contributions from phonons etc, and this should be 
taken properly into account 
in all the expressions above. However, this falls beyond our scope here. 

\section{Conclusions}

In this article we have presented arguments supporting 
the r\^ole of non-perturbative effects (instantons) 
in the non-Abelian $SU(2) \otimes U_S(1)$ model for high-temperature 
superconductors, discussed in \cite{fm},  
on the appearance 
of a `pseudogap ' phase, lying between the antiferromagnetic and 
the $d$-wave superconducting phases (c.f. fig. \ref{pseudogap}). 
The instantons have been argued to give a 
small mass to the gauge boson of the $U(1) \in SU(2)$,
which otherwise would have played the r\^ole of the massless pole
of the superconducting phase.  The Kosterlitz-Thouless nature of the 
symmetry breaking~\cite{dor,fm} was held responsible for the 
absence of phase coherence, which characterises the pseudogap phase.

The basic feature of the model is the relativistic nature 
of the charged excitations, relevant for the 
description of quasiparticles about the four points 
constituting 
the (putative) fermi surface in the pseudogap phase~\cite{fmb,fisher}. 

We have also argued that above a given doping concentration, 
higher than the ones at which gauge interactions set in,
four-fermi contact interactions become relevant, and they induce
a gap in the quasiparticle spectrum. For simplicity we have considered 
Gross-Neveu type interactions, although the 
analysis can be extended to other types.  

We have pointed out that, 
experimentally, such interactions may be distinguished from 
the gauge interaction by observing a change in the scaling properties
of the thermal conductivity of the nodal quasiparticles with an externally
applied magnetic field, for fixed temperatures, as the 
doping concentration changes. We have suggested the extension of 
the experiments of ref. \cite{ong} in the region of the 
pseudogap phase as well, because it is in this region that 
the gauge interactions have been argued to be relevant for the opening of 
a Kosterlitz-Thouless gap for the nodal excitations. 
We should note at this point that if the 
magnetic catalysis phenomenon occurs in the pseudogap phase, in much the same way as it occurs in the superconducting case of ref. \cite{ong}, 
then, according to the discussion in this article, there is a significant 
possibility that the dynamics of this phase is dominated by 
{\it electrically charged} nodal excitations, similar to 
the ones about the nodes of the $d$-wave superconducting 
gap suggested in \cite{fmb}, contrary to current 
theoretical scenaria~\cite{fisher}.

Before closing we would like to remark that 
an alternative way of probing the gauge structure 
in doped antiferromagnetic materials 
has been 
discussed in ref. \cite{momen}. The presence of gauge interactions 
leads to induced parity-violating magnetic moments
in effective low-energy theory models of high-temperature 
superconductors, 
under the influence of strong 
external magnetic fields, after integrating out fermionic degrees of freedom
pertaining to higher Landau levels. These magnetic moments are induced 
even in the case where the nodal gap is 
{\it parity conserving}. 
As shown in \cite{momen}, 
the scaling of the magnetic moment in the effective theory 
for the lowest Landau level, 
induced by the massive $SU(2)$ gauge bosons of 
the model of \cite{fm} in the gapped phase, 
is different 
from the scaling 
in the case where the moment 
is induced by the $U_S(1)$ interactions, as well as 
from the corresponding case of 
parity-violating Chern-Simons theories~\cite{szabo}.
These are experimentally testable predictions, which 
provide independent probes of gauge interactions, that are complementary 
to the 
thermal conductivity method described here and in refs. \cite{fmb,ssw}.

We are well aware that 
our effective theory analysis may be too crude to allow for a quantitative
comparison with the realistic situations, but we believe 
that the proposed phase structure, and the associated non-trivial 
symmetry 
breaking mechanisms that characterise the model of \cite{fm}, 
may capture  
essential features of the phase structure 
of high-temperature superconductivity,
if the above scenaria are  
realised in nature. 

\section*{Acknowledgements} 

The authors would like to thank I.J.R. Aitchison and J. Betouras 
for discussions. 
K.F. wishes to thank the TMR project FMRX-CT97-0122 for partial 
financial support, and the Department of Theoretical Physics 
of Oxford University for the hospitality during the last stages of this work. 
G.K. wishes to acknowledge partial financial support from 
PENED 95 Program No. 1170 of the Greek 
General Secretariat of Research and Technology.


\end{document}